# Low-field dephasing time of an optically polarized $NV^-$ spin bath beyond the P1 limit in diamond: a cluster-correlation-expansion and exact spin-bath study

Chikara Shinei[1,2,*]

[1] Department of Applied Physics, Institute of Pure and Applied Sciences, University of Tsukuba, Tsukuba, Ibaraki 305-8573, Japan

[2] Japanese-French Laboratory for Semiconductor Physics and Technology J-FAST, CNRS, University Grenoble Alpes, Grenoble INP, University of Tsukuba, Tsukuba, Japan

[*] e-mail: shinei.chikara.fb@u.tsukuba.ac.jp

## Abstract

Dense ensembles of nitrogen-vacancy ($NV^-$) centres in nitrogen-rich diamond dephase on the electron spins of the substitutional nitrogen (P1) centres. If the paramagnetic nitrogen is instead converted into $NV^-$ centres, the laser that initializes the sensor also pumps the bath into $m_s = 0$, a state with no moment along its own axis. Using cluster-correlation-expansion (CCE) calculations, combined with an exact solver for probe–bath exchange and with no adjustable parameter, this paper asks whether such an optically polarized $NV^-$ bath dephases a sensor more slowly than a P1 bath of the same density. The answer is set by the field. In a field along [111], three quarters of the bath centres lie off the field axis, and the moment of their $m_s = 0$ state along the field grows from 0.04 at 25 G to 0.46 at 300 G and 0.81 at 800 G. At 25 G the polarized bath is therefore nearly silent: at $10^{17}$ $cm^{-3}$ the dephasing time $T_2^*$ of a separately addressed $^{15}NV^-$ probe is 112 μs, six times the 18.1 μs of a P1 bath, and still 2.4 times at 90 % polarization; an ensemble measurement gains a factor of two. At 300 G the advantage over P1 shrinks to 1.4, and at 800 G the polarized bath dephases faster than P1. The field dependence of every bath, P1 included, follows the mean moment of its centres along the field. For the Hahn echo, exchange with on-axis bath centres and bath relaxation ($T_1$ = 6 ms) limit the gain to 1.2 times the P1 value at 25 G. Optical polarization of an $NV^-$ bath at low field therefore offers dephasing times beyond the P1 limit of nitrogen-rich diamond.

## Introduction

The nitrogen-vacancy centre ($NV^-$) in diamond is the most widely used solid-state spin sensor. Its spin can be initialized optically into $m_s = 0$ and read out at room temperature.[1] Sensitivity improves with the number of sensing spins, so magnetometers use dense ensembles. Density, however, brings the sensor close to other electron spins. The Hahn-echo coherence time $T_2$ sets the ac sensitivity and the free-induction-decay (FID) dephasing time $T_2^*$ sets the dc sensitivity; both shorten as the paramagnetic concentration rises.[2] In nitrogen-rich diamond the dominant bath is the substitutional nitrogen centre P1. For P1 both times scale as the inverse of the nitrogen concentration,[3] and cluster-correlation-expansion (CCE) calculations reproduce the measured scaling without adjustable parameters.[4] The mechanism is spectral diffusion. Pairs of bath spins exchange spin projections (flip-flop), and the dipolar field they produce at the sensor fluctuates. A flip-flop conserves energy only if both partners share an electron-paramagnetic-resonance (EPR) transition frequency. Anything that spreads those frequencies slows the bath down, as a companion study showed for multi-vacancy baths.[5]

Converting nitrogen into $NV^-$ centres changes the bath in a second way. In a high-density $NV^-$ ensemble a large part of the paramagnetic population is $NV^-$ itself, and $NV^-$–$NV^-$ interactions already govern the depolarization and thermalization of such ensembles.[6],[7] Unlike P1, an $NV^-$ bath can be polarized. The green laser that initializes the sensor also pumps every bath $NV^-$ into $m_s = 0$, a state with no magnetic moment along its axis. A fully polarized $NV^-$ bath has no flippable pairs, and in $m_s = 0$ its static field at the sensor vanishes to first order. Optical polarization should therefore lengthen both $T_2$ and $T_2^*$. The question this paper asks is whether a nitrogen-rich diamond in which the paramagnetic nitrogen has been converted into $NV^-$ centres and optically polarized can dephase a sensor more slowly than the same diamond limited by P1, the usual situation.

Three effects complicate the simple picture. First, an $NV^-$ centre whose axis is not parallel to the field is not a pure $m_s$ = 0 state. In a field along [111], three of the four $NV^-$ orientations sit at 109.5° to the field, and the transverse component mixes $m_s = \pm1$ into their ground state. The same transverse field degrades their optical polarization.[8,9] Second, bath $NV^-$ centres on the sensor's own axis share its transition frequency. A polarized bath therefore offers resonant partners into which the sensor's excitation can hop. Third, the bath polarization is not permanent. It relaxes back to equilibrium with the longitudinal time $T_1$ of the bath centres, a few milliseconds at room temperature, and every random jump moves a dipolar field.

These effects depend on the magnetic field and on how the sensor is addressed, and each pulls in a different direction. This paper treats all three at 25, 300 and 800 G along [111]. These fields span weak, intermediate and strong mixing of the off-axis centres, below the ground-state level anticrossing at 1024 G. Three addressing schemes are compared. (i) A probe $NV^-$ that is the only spin driven. (ii) A $^{15}NV^-$ probe in a $^{14}NV^-$ bath. Its $^{15}N$ hyperfine lines lie between the $^{14}N$ lines, so selective pulses address it alone; such probes are made by $^{15}N$ implantation or δ-doping.[10,11] (iii) An ensemble measurement that drives every $NV^-$ of one orientation. A P1 bath of the same density, computed with the same methods, is the benchmark throughout. The calculation uses CCE[12,13] for bath flip-flops, an exact excitation-number-conserving solver for probe–bath exchange, and a Markov model for $T_1$. Its only inputs are the $NV^-$ and P1 spin-Hamiltonian parameters, including the $^{14}N$ hyperfine splitting of $NV^-$.[14]

---

## Results

### Level structure and the induced moment of off-axis centres

Figure 1a shows the single-centre levels in a field along [111]. For the on-axis orientation, the qubit transition $m_s = 0 \leftrightarrow -1$ lies at 2799.9, 2029.05 and 627.5 MHz at 25, 300 and 800 G for $m_I = 0$. The $^{14}N$ hyperfine shifts it by ±2.16 MHz for $m_I = \pm1$. The transitions of the off-axis centres lie at least 49 MHz away from the probe at every field (2.85 and 2.90 GHz at 25 G, 2.89 and 3.47 GHz at 300 G, 1.71, 3.74 and 5.45 GHz at 800 G). The on-axis centres with the same $m_I$ as the probe, one bath centre in twelve, are therefore the only exact resonances of the probe.

Figure 1b shows the moment of the off-axis $m_s = 0$ eigenstate. Along the field it is −0.043 at 25 G, −0.46 at 300 G and −0.81 at 800 G, with full moments $|\langle \mathbf{S} \rangle|$ of 0.046, 0.49 and 0.84. At low field the transverse mixing enters at second order, and a polarized $NV^-$ bath becomes nearly silent. At 300 G three quarters of the bath keep about half the moment of an unpolarized spin even when fully polarized. At 800 G they keep almost all of it, and every off-axis moment points the same way. This field dependence controls most of what follows.

### Dephasing: $T_2^*$

Figure 2b,d,f shows the FID at $10^{17}$ $cm^{-3}$ and Fig. 3a–c the concentration dependence (Table 1; Supplementary Table 1). The benchmark is a P1 bath, with $T_2^*$ = 18.1, 11.6 and 11.5 μs at 25, 300 and 800 G. An unpolarized $NV^-$ bath dephases the probe slightly faster, 10.1, 9.3 and 7.9 μs, because $NV^-$ has $S = 1$ and the larger mean moment.

For a $^{15}NV^-$ probe (see below), full polarization ($p = 1$, where $p$ is the population of $m_s = 0$) lengthens $T_2^*$ from 10.4 to 112 μs at 25 G. This is six times the P1 value of 18.1 μs: at low field a polarized $NV^-$ bath dephases the sensor far more slowly than the P1 bath it replaces. The advantage survives incomplete polarization, 43 μs or 2.4 times P1 at $p = 0.9$, and an ensemble measurement, 35 μs or 1.9 times P1 (see Ensemble measurements). It rests on the off-axis centres: polarizing only the on-axis quarter of the bath gives 13.8 μs, below P1. At $10^{15}$–$10^{16}$ $cm^{-3}$ the polarized bath at 25 G brings $T_2^*$ close to the $^{13}C$ limit of about 400 μs. At 300 G the gain shrinks to 1.6, from 9.8 to 15.8 μs, and the advantage over P1 to 1.4. At 800 G it nearly vanishes, 8.4 to 9.1 μs, and the polarized $NV^-$ bath falls below a P1 bath (11.5 μs). At 800 G the polarization of the off-axis centres is even counter-productive. Leaving them unpolarized while the on-axis centres are polarized gives 10.7 μs, longer than full polarization, because a fully polarized bath aligns all off-axis moments.

The field dependence of all three baths has one origin. In the static limit a dilute dipolar bath gives a Lorentzian decay whose width is proportional to the mean moment of the bath centres along the field, so that $T_2^* \propto 1/\langle|\langle S_z \rangle|\rangle$.[16] For P1 this mean is 0.30 at 25 G and 0.50 at 300 and 800 G. At 25 G the electron Zeeman energy (70 MHz) is smaller than the $^{14}N$ hyperfine coupling (81–114 MHz), the electron is quantized partly along its hyperfine axis rather than the field, and its moment along the field shrinks. The ratio predicts a 1.65-fold longer $T_2^*$ at 25 G than at 300 G; the computed ratio is 1.56. For an unpolarized $NV^-$ bath the mean is 0.34, 0.46 and 0.59: at low field the off-axis centres are quantized along their own axes, at 109.5° to the field, and their $m_s = \pm1$ moments project onto the field by about one third; at high field they turn towards the field. For the polarized bath the mean falls to 0.033 at 25 G, against 0.35 and 0.61 at 300 and 800 G. The moment argument ignores the transverse components of $\langle \mathbf{S} \rangle$ and the dispersive exchange shifts, which the calculation includes, and it overestimates the low-field gain of the $NV^-$ baths (Supplementary Note 8), but it captures every trend.

At 25 G the polarized $NV^-$ bath carries a mean moment nine times smaller than P1, and this is why it outperforms P1 there.

Figure 4a shows the dependence on $p$. At 25 G $T_2^*$ rises steeply above $p \approx 0.9$, because residual $m_s = \pm 1$ centres then dominate the static field; it is 43 μs at $p = 0.9$ and 112 μs at $p = 1$. At 300 G it rises slowly to 1.6 times, and at 800 G it is flat. These values come from the exact solver, which adds the dispersive shifts $t^2/\Delta$ of off-resonant exchange to the static fields. For the $^{15}NV^-$ probe these shifts shorten $T_2^*$ by at most 10–15 % relative to the static-field model. For a $^{14}NV^-$ probe at 25 G, resonant exchange halves it, to 66 μs. $T_2^*$ is therefore set mainly by static fields, and by exchange only when the probe shares the bath transition.

**Probe–bath exchange and its localization**

In the Hahn echo, polarization removes bath flip-flops. In conventional CCE with the bath at $p = 1$, the echo does not decay at all. What remains is exchange. The probe component in $m_s = -1$ can hop to a resonant on-axis centre in $m_s = 0$, and the ideal π pulse acts on the probe only. The exact solver gives this "exchange factor" (Fig. 2a,c,e, dotted, and Supplementary Fig. 2). It falls by a fraction and then stays on a plateau. At $2\tau = 10$ ms and $10^{17}$ cm$^{-3}$ the plateau is 0.41 at 25 G, 0.74 at 300 G and 0.83 at 800 G (Fig. 5c).

The plateau and its field dependence come from localization. The excitation can move only through the on-axis $m_s = 0$ sublattice, which is dilute and at random positions. Its sites are detuned from one another by the static fields of all the other centres. As the field grows, the induced moments of the off-axis centres make that disorder larger than the hopping, and the excitation stays near the probe. Removing the off-axis static field in the model drops the plateau to 0.16 at every field (Fig. 5c). The $^{14}N$ hyperfine also helps. It confines resonance to one $m_I$ in three, and removing it lowers the plateau to 0.20, 0.47 and 0.58 at the three fields.

The field therefore acts in opposite directions on the two coherence times of a $^{14}NV^-$ probe. The moment that limits $T_2^*$ at high field is the same one that localizes exchange and preserves the echo.

**Field dependence**

Figure 5 collects the three fields at $10^{17}$ cm$^{-3}$. The $T_2^*$ gain from polarization falls monotonically with field: 10.8, 1.6 and 1.08 for the $^{15}NV^-$ probe, and 3.3, 1.3 and 0.95 for an ensemble measurement (Fig. 5b). It tracks the off-axis moment, which rises from 0.04 to 0.81 over the same range. The exchange plateau rises monotonically with field (Fig. 5c), and so does $T_2$ with $T_1$ included (Fig. 5d, discussed below). The induced moment of the off-axis centres is thus the single quantity that decides these outcomes. A low field makes the polarized bath magnetically silent. A high field makes it a strong static disorder that blocks exchange and flip-flops but dephases the probe.

**A spectrally distinct probe and longitudinal relaxation**

The exchange channel exists only because the probe and the bath share a transition. A $^{15}NV^-$ probe has its $0 \leftrightarrow -1$ transition shifted by ±1.515 MHz, 0.645 MHz from the nearest $^{14}N$ line, far beyond any dipolar coupling. The exact solver confirms that the exchange factor stays above 0.95 to 10 ms at all concentrations and fields. Without $T_1$, the echo is then limited by flip-flops of residual $m_s = \pm 1$ centres and by instantaneous diffusion within the $^{15}NV^-$ group itself. For a $^{15}NV^-$ fraction of 1 %, $T_2$ is 10 ms at $10^{17}$ cm$^{-3}$ at every field (Fig. 2a,c,e, dashed).

Longitudinal relaxation changes that number substantially. Three effects of a room-temperature $T_1$ of 6 ms are included (Methods). The probe loses coherence as $\exp(-t/2T_1)$. The bath polarization decays towards 1/3 during the sequence. Each bath centre also jumps at random between its levels. Jumps are not refocused by the echo, and their effect grows with the static field each jump moves, which scales as $1/T_2^*$. This is the classic sudden-jump spectral diffusion.[18] Alone, it limits the echo to 0.42–0.50 ms at $10^{17}$ cm$^{-3}$, depending on the field, and it scales as $n^{-1/2}$: 5.4, 1.5, 0.47 and 0.15 ms from $10^{15}$ to $10^{18}$ cm$^{-3}$ at 300 G (Supplementary Fig. 4). The decay of the polarization adds to it, because every relaxed centre is a minority $m_s = \pm 1$ spin that can flip-flop. How fast those minority centres flip-flop depends strongly on the field. Without $T_1$, $T_2$ of the $^{15}NV^-$ probe at $p = 0.9$ is 121, 208 and 663 μs at 25, 300 and 800 G (Supplementary Table 3). At high field the induced moments of the polarized off-axis centres detune the minority pairs, the same static disorder that localizes exchange.

With $T_1$ included, the $^{15}NV^-$ probe in a fully polarized bath reaches $T_2$ = 260, 358 and 376 μs at 25, 300 and 800 G at $10^{17}$ cm$^{-3}$ (Fig. 3d–f; Fig. 5d; Table 2; Supplementary Table 2). These values are 3.0, 3.9 and 2.3 times the unpolarized $NV^-$ bath (85, 91 and 162 μs). At 300 and 800 G they are 88 and 92 % of a P1 bath of the same density (408 μs). At 25 G the P1 bath itself is shorter, 212 μs, because the electron–nuclear mixing of P1 weakens the mean field that slows its flip-flops, and the polarized $NV^-$ bath exceeds it by a factor 1.2. Unlike $T_2^*$, $T_2$ grows with the field, because the minority flip-flops that follow the polarization decay are slower at high field. The gain from polarization grows with concentration, because flip-flops scale as $1/n$ and jumps only as $n^{-1/3}$. At $10^{18}$ $cm^{-3}$ it is 6.6–12 times (74–115 against 8.5–17 μs), and the polarized bath then exceeds a P1 bath (42 μs at 300 and 800 G, 22 μs at 25 G) by a factor of 2.6–3.4. At $10^{16}$ $cm^{-3}$ the gain is 1.2–1.5 times and P1 is three times better, and at $10^{15}$ $cm^{-3}$ $T_2$ is about 4 ms with or without polarization, because the polarization is itself lost on the $T_1$ timescale. With $T_1$ included, $T_2$ depends strongly on the initial polarization (Fig. 4b). At $p = 0.9$ it is 115, 165 and 282 μs, 44–75 % of full polarization, and the loss is largest at low field. Most of the benefit therefore requires polarization close to unity, especially at low field.

**Ensemble measurements**

In an ensemble measurement the microwave pulses drive every $NV^-$ of the probed orientation, a quarter of all $NV^-$, together. Those centres are then not a polarized bath but other probes in superposition. Their mutual Ising coupling is not refocused by the echo (instantaneous diffusion).[15] The calculated decay reproduces the analytic rate $1/T_{ID} = 8\pi^3 \mathcal{J}_0 n/(9\sqrt{3})$, with $n$ the density of driven centres, to within 3 %. It gives $T_2 \approx 50$ μs at $10^{17}$ $cm^{-3}$ at every field, whether the remaining three orientations are polarized or not. This is shorter than the P1 bath (Fig. 3d–f). Polarization lengthens the ensemble $T_2^*$ only at low field: from 10.7 to 35 μs at 25 G and from 9.4 to 12.4 μs at 300 G, while at 800 G it shortens slightly, from 8.8 to 8.4 μs (Fig. 3a–c). In the ensemble geometry polarization therefore helps mainly the dc sensitivity, and only at low field. Lengthening $T_2$ further requires suppressing the like-spin interaction, for example by dipolar decoupling sequences.[19,20]

---

## Discussion

The hypothesis that an optically polarized $NV^-$ bath protects the sensor holds, but with three qualifications. Each traces to one physical fact.

First, polarization is magnetically complete only at low field. A centre in $m_s = 0$ carries no moment only when the field lies along its axis. In a field along [111], three quarters of the bath sit at 109.5°, and their induced moment grows from 0.04 at 25 G to 0.46 at 300 G and 0.81 at 800 G. The $T_2^*$ gain falls accordingly, from an order of magnitude to essentially none. At 800 G, polarizing the off-axis centres even shortens $T_2^*$, because it aligns their moments. Choosing the bias field is therefore part of the bath engineering. For $T_2^*$, the lowest field compatible with resolving the $NV^-$ orientations is best. A low field also makes the off-axis centres easier to polarize optically.[8,9] At 800 G, where the transverse field is 754 G, their optical polarization would be poor in any case.

Second, a polarized bath of identical centres is not inert towards an identical probe. It offers resonant partners for the probe's own excitation. The same static disorder that limits $T_2^*$ at high field localizes that exchange, so the echo of a $^{14}NV^-$ probe is better preserved at 800 G than at 25 G. A $^{15}NV^-$ probe avoids the problem altogether. For it, the field choice is a trade-off: $T_2^*$ favours low field, whereas $T_2$ at room temperature favours high field, where the same disorder slows the minority flip-flops that follow bath relaxation (260 μs at 25 G against 376 μs at 800 G at $10^{17}$ $cm^{-3}$). A practical layout is a $^{15}NV^-$-doped sensing layer in a $^{14}N$-grown diamond, made by implantation or δ-doping.[10,11] It then needs pulses selective to within about 0.6 MHz.

Third, at room temperature the bath $T_1$ decides the ceiling. Random jumps of bath centres alone limit $T_2$ to about 0.45 ms at $10^{17}$ $cm^{-3}$, and the minority flip-flops created by relaxation lower it to 0.26–0.38 ms. That is far below the $2T_1$ = 12 ms bound and the $^{13}C$ limit of 200 ms at 50 ppm. At this density a polarized $NV^-$ bath is therefore not better than a P1 bath of the same density at 300 and 800 G; it becomes better above about $1.5 \times 10^{17}$ $cm^{-3}$, where P1 flip-flops ($\propto 1/n$) outpace the jumps ($\propto n^{-1/3}$). At 25 G it is already 1.2 times better at $10^{17}$ $cm^{-3}$, because the low field also shortens the P1 echo. This bound applies to any paramagnetic bath that relaxes, P1 included, so P1 values with their own $T_1$ jumps would also be lower than shown. $NV^-$–$NV^-$ cross-relaxation shortens the effective $T_1$ further in dense ensembles,[6] and the polarized-bath gains reported here are then upper bounds. Lowering the temperature lengthens $T_1$ steeply.[17] The polarized-bath $T_2$ would then approach the 10 ms set by the residual flip-flops and the probe group.

Two technical findings matter beyond this system. First, the generalized CCE as implemented in PyCCE applies ideal pulses as instantaneous rotations in the laboratory frame. When the probe exchanges resonantly with its bath, the two echo pathways then acquire a spurious relative phase $\exp(2i\omega_0\tau)$, which a real microwave pulse does not produce. Grids commensurate with the probe frequency remove it (Methods; Supplementary Fig. 1). Second, the CCE of an electron-spin bath converges only if the bath states carry the static mean field of the spins outside each cluster, and the cluster cutoff is several mean spacings. With a fully mixed bath, or with a cutoff near one spacing, the echo time of an unpolarized P1 or $NV^-$ bath is underestimated by up to a factor of four, and a pure-state sampled bath appears frozen. This agrees with the convergence analysis of Park et al.[24] All CCE results here use pure-state sampling with a cutoff of four mean spacings (Methods). Both effects apply to any calculation of a qubit embedded in an electron-spin bath.

Several limitations remain:

- The minority $m_s = \pm 1$ centres are frozen in the exchange solver and treated to second order in CCE, with the mean field of the other centres.
- The concentration dependence of the flip-flop factor is obtained by scaling a calculation at $10^{17}$ cm$^{-3}$ (Methods).
- The polarization decay uses a quasi-static average over the sequence.
- P1 $T_1$ is not included.
- The absolute $T_2^*$ of the P1 bath, 6.9 µs·ppm at 300 G, is about 30 % below the experimental 9.6 µs·ppm.[3] The static-bath theory behind it is exact for a dilute dipolar bath, and second-order CCE changes it by only 2 %. The difference probably reflects how the P1 concentration is determined experimentally. It applies to all baths alike and leaves the ratios unchanged.

In summary, at low field an optically polarized NV$^-$ bath gives a dephasing time six times longer than a P1 bath of the same density, 2.4 times at 90 % polarization and twice in an ensemble measurement, because the polarized off-axis centres carry almost no moment along the field. The advantage shrinks as the field grows and reverses at 800 G. At room temperature it extends $T_2$ by 2–4 times at $10^{17}$ cm$^{-3}$ and up to 12 times at $10^{18}$ cm$^{-3}$, but only for a probe that the measurement can address separately from the bath and only for polarization close to unity. In practice not all paramagnetic nitrogen becomes NV$^-$, and the bath is not fully polarized. Fig. 6 shows what an experiment can expect as a function of both. It maps the ratio of $T_2^*$ to that of a P1-only bath of the same total density when a fraction $f$ of the paramagnetic nitrogen is NV$^-$ with polarization $p$ and the rest is P1 (Methods). At 25 G a twofold gain requires $f \gtrsim 0.6$ at $p = 1$ or $f \gtrsim 0.85$ at $p = 0.9$, and any gain at all requires $p$ above 0.68, where a polarized NV$^-$ bath first dephases more slowly than P1. At $f = 0.5$ the gain is 1.7 at $p = 1$ and 1.4 at $p = 0.9$; at $f = 0.3$ it is 1.3. At 300 G the gain never exceeds 1.4. An ensemble measurement gains less, because the mutual coupling of the driven centres is not removed by polarization (Fig. 6c,d): at 25 G the ratio is 1.95 at $f = 1$ and $p = 1$, 1.5 at $f = 0.7$ and 1.3 at $f = 0.5$, and it exceeds unity only above $p = 0.72$; at 300 G it stays below 1.07. The figure therefore sets the materials targets, a high conversion of paramagnetic nitrogen into NV$^-$ and a polarization above about 0.9, and shows that both pay off only at low field.

These results give quantitative design rules for dense NV$^-$ ensembles:

- a low bias field for $T_2^*$, a higher one for $T_2$;
- polarization close to unity and a spectrally distinct probe species for $T_2$;
- dipolar decoupling when the ensemble itself is the sensor.

---

## Methods

### Spin Hamiltonian

The probe is an NV$^-$ centre with $S = 1$, zero-field splitting $D = 2.87$ GHz, $g = 2.0028$ and axis along [111], parallel to the field. Its qubit is formed by $m_s = 0$ and $-1$. Bath NV$^-$ centres are placed uniformly at random, the four ⟨111⟩ orientations and the three $^{14}$N projections $m_I$ being equally likely. Each bath centre has the spin Hamiltonian $D(\mathbf{n}\cdot\mathbf{S})^2 + g\mu_B\mathbf{B}\cdot\mathbf{S}$ with its own axis $\mathbf{n}$, diagonalized exactly, and the secular $^{14}$N hyperfine term $A_\| m_I(\mathbf{n}\cdot\mathbf{S})$, with $A_\| = -2.16$ MHz.[14] The transverse hyperfine term is suppressed by $D \pm \gamma_e B$ and shifts levels by a few kHz; it is neglected. All spins interact through the full point-dipole coupling. The probe has $m_I = 0$, a choice that does not affect the results because the bath $m_I$ are uniform. The three fields stay clear of the level anticrossings: NV$^-$–P1 cross-relaxation near 512 G and the ground-state anticrossing at 1024 G.

P1 centres ($S = 1/2$, $g = 2.0024$) carry the $^{14}$N hyperfine tensor ($A_\| = 114.03$ MHz, $A_\perp = 81.31$ MHz) along one of four Jahn–Teller axes, and a quadrupole term of −3.97 MHz.[21,22] For bath dynamics at 300 and 800 G the hyperfine is reduced to a scalar detuning, as in the companion study.[5] In that limit the P1 bath dynamics do not depend on the field. For static fields the electron–$^{14}$N problem is diagonalized exactly at all fields. At 25 G the electron Zeeman energy (70 MHz) is smaller than the hyperfine and the high-field reduction fails. The P1 echo at 25 G is therefore taken from a calculation that keeps the nitrogen nucleus explicit: each P1 is an electron and its $^{14}$N coupled by the full hyperfine and quadrupole tensors, the cluster expansion runs in units of whole centres, each cluster is kept fully mixed, and the static mean field of the outer centres is sampled from their electron–nuclear eigenstates (the product-state sampling of the next subsection cannot represent these entangled states). It uses the same probe, field and cutoff of four mean spacings, 24 configurations at 1 ppm, and the same density scaling ($T_2 n$ = 123 μs·ppm). For P1 at 300 G this mean-field sampling reproduces full pure-state sampling within the statistical uncertainty.

### Bath states and polarization

Each bath $NV^-$ is described in the eigenbasis of its single-centre Hamiltonian. The $m_s$ = 0-like eigenstate is the one that minimizes $\langle(\mathbf{n}\cdot\mathbf{S})^2\rangle$. A bath of polarization $p$ has each centre in its $m_s$ = 0-like eigenstate with probability $p$ and in each of the other two with probability $(1 - p)/2$; $p = 1/3$ reproduces the infinite-temperature bath. The static, exchange and CCE calculations draw the eigenstates from these populations for each configuration. On-axis and off-axis orientations can be given different $p$.

### Bath flip-flops: conventional CCE

The echo decay caused by flip-flops among bath spins is computed by second-order conventional CCE (PyCCE 1.1.1).[13] The bath state is sampled as a pure product state: each centre is assigned one eigenstate of its single-centre Hamiltonian, drawn with the populations of the bath ($p$, $(1 - p)/2$, $(1 - p)/2$ for $NV^-$; equal for P1). Spins outside each cluster enter through the static mean field of their sampled states. Each spatial configuration is averaged over four state samples (one at $p = 1$), and the complex coherence is averaged over 12 configurations. The cluster cutoff is 4 $n^{-1/3}$ and the bath sphere has a radius of 5.5 $n^{-1/3}$ (about 700 centres). This choice follows a convergence test (Supplementary Note 7): with a fully mixed bath the echo time of an unpolarized bath converges to a value up to four times too short, and with pure-state sampling a cutoff near one mean spacing freezes the bath. With the mean field and a cutoff of 3–5 spacings, the P1 echo converges to $T_2 n$ = 214–282 μs·ppm (236 μs·ppm in the production run), in agreement with the 236–286 μs·ppm of Park et al.[24] Because only the dipolar coupling scales with density, and the hyperfine and orientational detunings exceed it by orders of magnitude, $T_2 n$ is independent of $n$ (to 0.1 % between $10^{16}$ and $10^{17}$ $cm^{-3}$ in the test). The flip-flop factor is therefore computed at $10^{17}$ $cm^{-3}$ and transferred to other densities as $L_n(t) = L_{10^{17}}(tn/10^{17}\ cm^{-3})$.

### Probe–bath exchange: exact solver

In the secular approximation the probe and the on-axis bath centres in $m_s$ = 0 form two-level systems ($|0\rangle$, $|-1\rangle$), and the Hamiltonian conserves the number of $|-1\rangle$ excitations:

$$H = \Sigma_k\, \varepsilon_k\, n_k + \Sigma_{i<k}\, t_{ik}(\sigma^+_i\sigma^-_k + \text{h.c.}) + \Sigma_{i<k}\, V_{ik}\, n_i\, n_k,$$

with $V_{ik} = \mathcal{J}_0(1 - 3\cos^2\theta_{ik})/r_{ik}^3$, $t_{ik} = -V_{ik}/2$ and $\mathcal{J}_0 = \mu_0 g^2\mu_B^2/(4\pi h)$ = 52.1 MHz $nm^3$. The factor 1/2 in $t$ follows from $\langle 0|S_+|-1\rangle = \sqrt{2}$. The on-site energies $\varepsilon_k$ contain the $^{14}N$ offset and the static dipolar field of every other centre. Those other centres are the off-axis centres and the on-axis centres not in $m_s$ = 0. They are frozen in their eigenstates, with their full moment vectors. The FID is the return amplitude in the one-excitation sector. The Hahn echo with an ideal π pulse on the probe also needs the two-excitation sector:

$$L(2\tau) = |a_p(\tau)|^2 + \Sigma_{k\neq p}\, [U_2(\tau)\mathbf{a}(\tau)]_{(p,k)}\, a_k(\tau)^*,$$

where $a_k(\tau)$ are the one-excitation amplitudes and $U_2$ the two-excitation propagator. A $^{15}NV^-$ probe is modelled by shifting the probe on-site energy by $A_\parallel(^{15}N)/2$ = 1.515 MHz. Echo results are averaged over 96 configurations (24 for the $^{15}NV^-$ probe and the controls), FID results over 480.

### Static fields and ensembles

For P1 and for the ensemble geometry, exchange with the probe plays no role. There the FID is computed with bath spins frozen in their eigenstates ($T_2^* \ll T_2$):

$$L(t) = \langle \Pi_j\, \Sigma_w\, p_{jw} \exp(-2\pi i\delta_{jw}t)\rangle,$$

where $\delta_{jw}$ is the secular dipolar shift from centre $j$ in eigenstate $w$, including the transverse components of $\langle\mathbf{S}\rangle$. The average is taken over 1500 configurations of 1000 spins. It uses the complex coherence, which is what an ensemble signal measures. This model reproduces the analytic Lorentzian width of a dilute bath to within 3 %.[16]

For Fig. 6 the paramagnetic nitrogen at a fixed total density $n$ is split into $NV^-$ (fraction $f$, polarization $p$) and P1 (fraction $1 - f$). In the static limit the free-induction decay of independent dilute baths is a product of Lorentzians, so their dephasing rates add, $1/T_2^*(f, p) = f/T^*_{2\,NV}(p) + (1 - f)/T^*_{2\,P1}$, with both times taken at the total density. $T^*_{2\,NV}(p)$ is the $^{15}NV^-$-probe value of Fig. 4a, interpolated in log $T_2^*$ between the computed polarizations. For the ensemble measurement it is the product of the driven-group factor and the static FID of the three off-axis orientations at polarization $p$, both of which scale with the $NV^-$ density and hence with $f$.

In ensemble measurements every on-axis $NV^-$ is driven. The echo is then limited by the Ising coupling within that group, $L = \langle \Pi_j \cos(2\pi V_j\tau)\rangle$, and the FID by the factor $\langle \Pi_j\, [1 + \exp(2\pi i V_j t)]/2\rangle$. The remaining three orientations are treated as a bath (conventional CCE for the echo, static model for the FID). The $^{15}NV^-$ group is treated in the same way at 1 % of the $^{14}NV^-$ density, with one hyperfine line driven.

**Longitudinal relaxation**

With a single-quantum rate Ω, the $m_s = 0$ population relaxes as $\exp(-3\Omega t)$, so $T_1 = 1/(3\Omega)$. A room-temperature value of 6 ms is used at all fields.[17] The probe coherence then decays as $\exp(-t/2T_1)$.

The bath polarization relaxes as $p(t) = 1/3 + (p_0 - 1/3)\exp(-t/T_1)$. The flip-flop and exchange factors are interpolated in $p$ to the average over the sequence, $\bar{p}(t) = 1/3 + (p_0 - 1/3)(T_1/t)(1 - \exp(-t/T_1))$.

Random jumps of bath centres are computed exactly for independent centres. Each centre is a three-state Markov process with transitions $0 \leftrightarrow \pm 1$, and the phase it imprints on the probe is evaluated with transfer matrices for the FID and echo time signs. In the limit $T_1 \longrightarrow \infty$ this reduces to the static model to $10^{-12}$.

The $^{13}C$ contribution (50 ppm, the residual level of $^{12}C$-enriched crystals[23]) is computed by conventional CCE (order 2 for the echo, order 1 for the FID) and multiplied in. Independent factors are multiplied. $T_2$ and $T_2^*$ are the 1/e times of the averaged coherence. For the echo, a running minimum is applied to the electron-bath factors to suppress cluster divergence. The $^{13}C$ factor, which shows ESEEM at 25 G, is replaced by its upper envelope through the revivals (Supplementary Fig. 3).

**Validation**

For few-spin systems the exact solver agrees with full-cluster generalized CCE to within $3 \times 10^{-3}$, provided the time grid is an integer multiple of the probe period. On an arbitrary grid PyCCE's laboratory-frame pulses introduce echo errors up to 0.77 when the probe exchanges resonantly with the bath (Supplementary Fig. 1). All generalized-CCE echoes in this work therefore use commensurate grids.

Generalized CCE was not used as the main method because at second order it diverges for these electron baths.

**Code availability**

The simulation code is available from the corresponding author on reasonable request. It comprises the bath construction, the exact excitation-number solver, the static-field and $T_1$ models, the PyCCE drivers and the analysis scripts. The underlying package, PyCCE,[13] is openly available.

---

## Data availability

The data supporting the findings of this study are available from the corresponding author on reasonable request. They comprise the simulated coherence decays behind every figure and table, at all three fields, every concentration and every polarization, on a common 241-point logarithmic time axis, and the coherence times extracted from them. All input parameters are published spin-Hamiltonian parameters given in Methods.

## Acknowledgements

This study was funded by the Japan Society for the Promotion of Science (JSPS) KAKENHI Grant Number 25K17636. The funder played no role in study design, data collection, analysis and interpretation of data, or the writing of this manuscript.

## Author contributions

C.S. conceived the study, performed the simulations, analysed the results and wrote the manuscript.

## Competing interests

The author declares no financial or non-financial competing interests.

---

## Figure legends

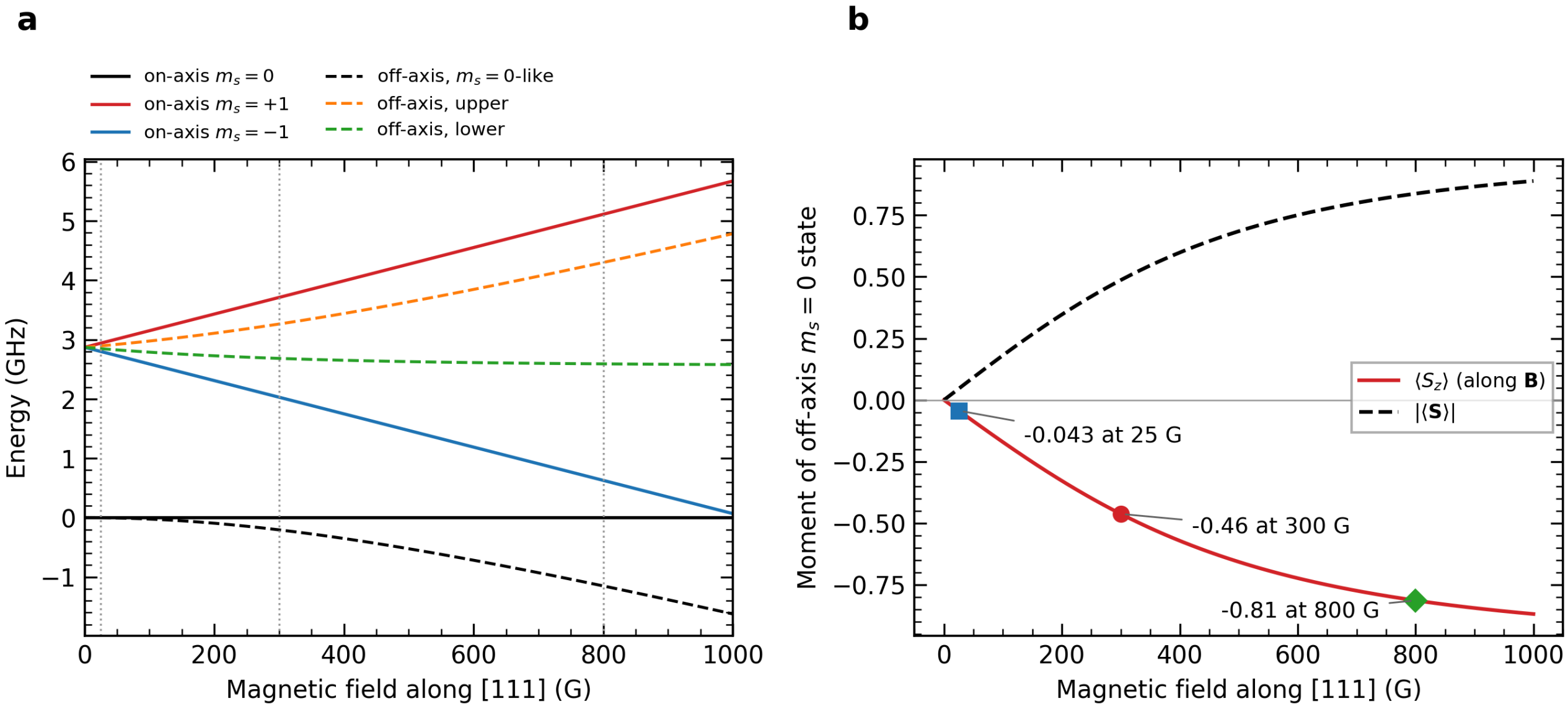

**Fig. 1 | NV⁻ levels in a field along [111] and the induced moment of off-axis centres. a**, Single-centre energy levels with $m_I = 0$, referenced to $m_s = 0$ at zero field ($m_s = \pm1$ at $D = 2.87$ GHz). Solid: on-axis orientation ($m_s = 0, +1, -1$). Dashed: an off-axis orientation at 109.5°, its $m_s = 0$-like level (black) and the upper and lower of the two remaining levels. Dotted vertical lines mark 25, 300 and 800 G. **b**, Moment of the off-axis $m_s = 0$-like eigenstate: component along the field $\langle S_z \rangle$ (red) and magnitude $|\langle \mathbf{S} \rangle|$ (dashed). It is −0.043, −0.46 and −0.81 at 25, 300 and 800 G, so a fully polarized bath is magnetically silent only at low field.

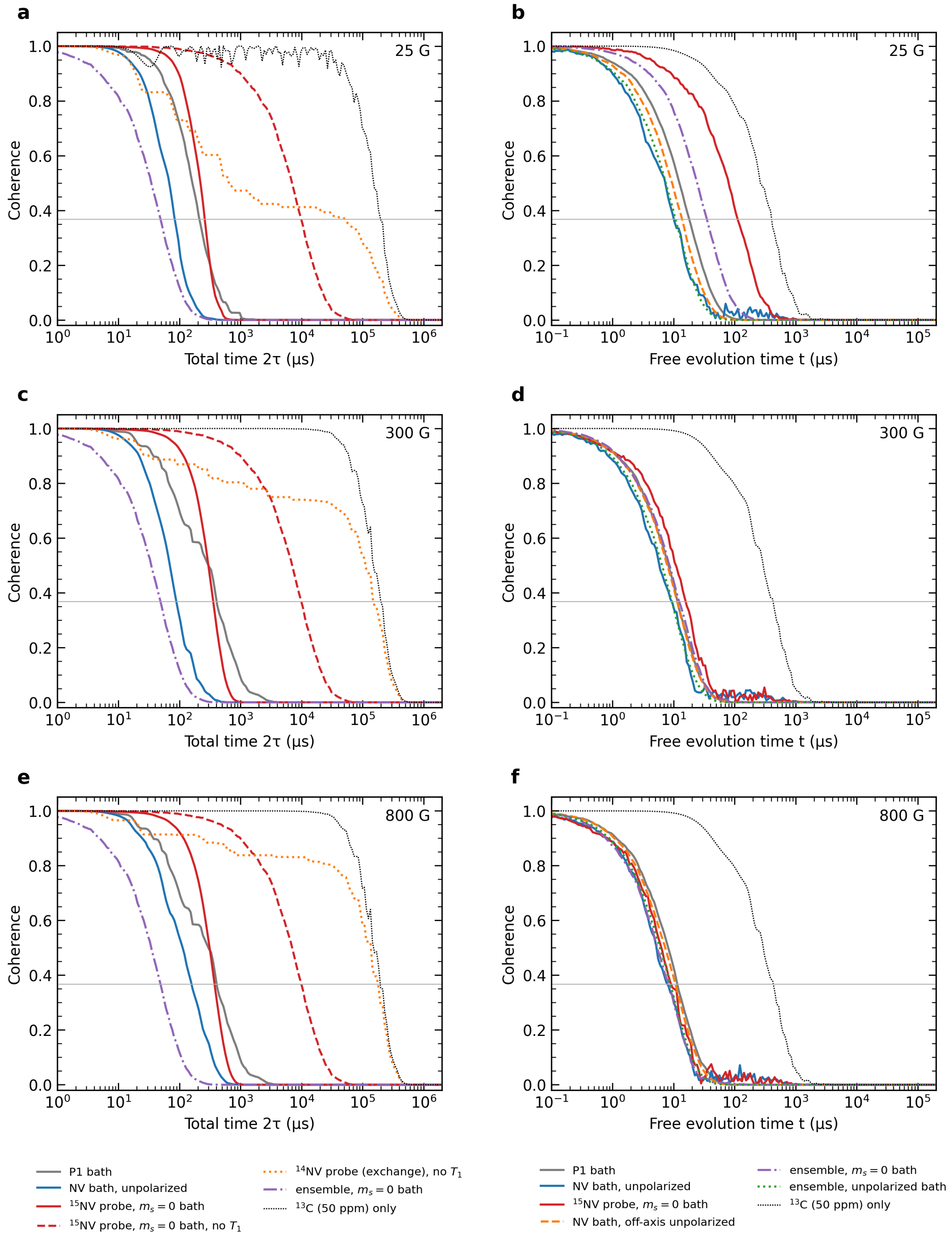

**Fig. 2 | Coherence decays at $10^{17}$ cm$^{-3}$. a**,**c**,**e**, Hahn echo at 25 G (**a**), 300 G (**c**) and 800 G (**e**). Solid lines include room-temperature $T_1$ = 6 ms; dashed and dotted lines omit it. The $^{14}NV^-$ probe (dotted orange) shows the exchange plateau in a fully polarized bath, which rises with field. A $^{15}NV^-$ probe (red) is off resonance with the bath and, without $T_1$, decays only at about 10 ms. **b**,**d**,**f**, Free induction decay at the same fields. The black dotted line is the $^{13}C$-only (50 ppm) reference; at 25 G it shows ESEEM. Horizontal grey lines mark 1/e. Legends below each column apply to all rows; at 25 G the P1 echo is computed with the nitrogen nucleus explicit (Methods).

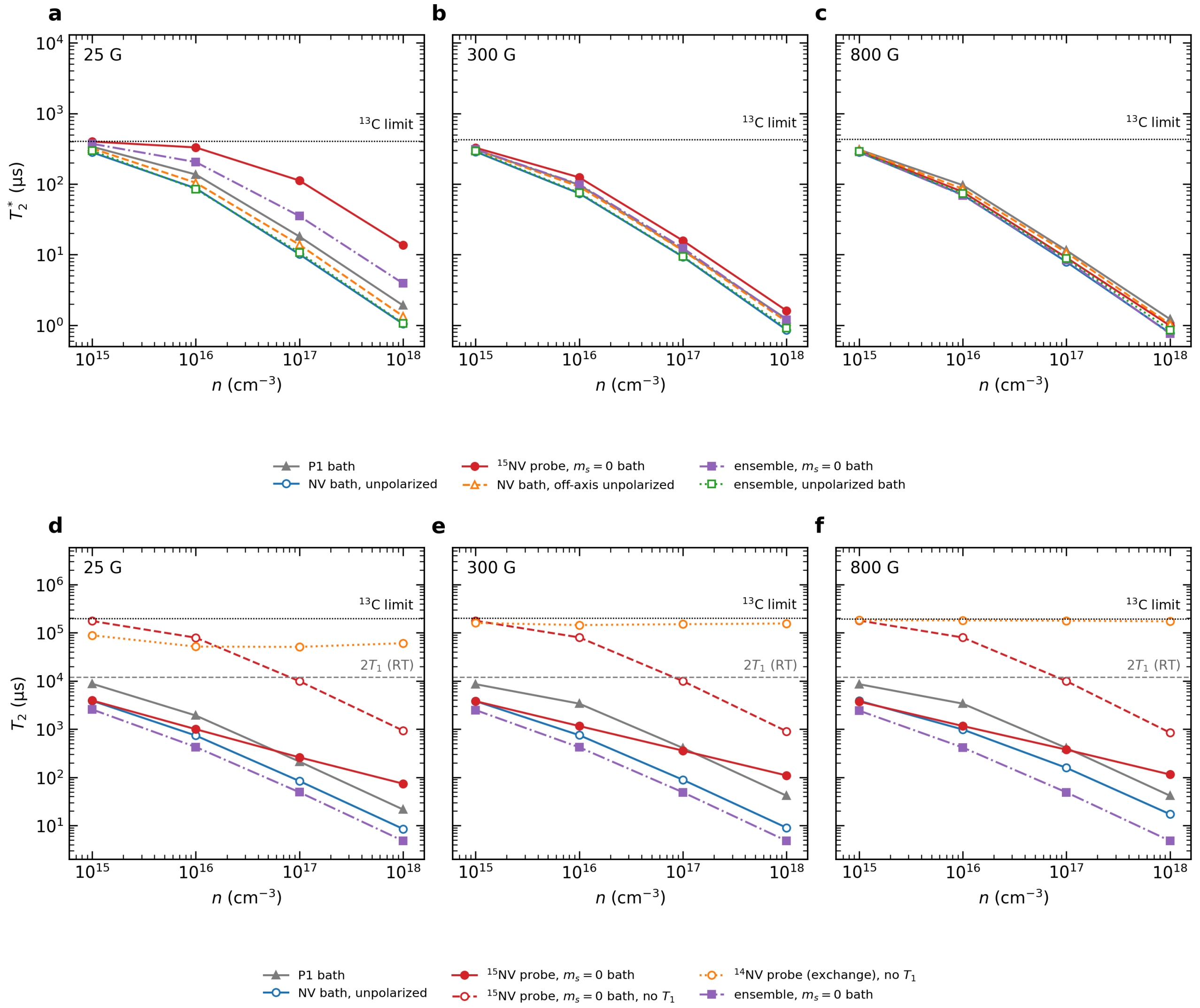

**Fig. 3 | Concentration dependence of $T_2^*$ and $T_2$. a–c**, $T_2^*$ at 25, 300 and 800 G. **d–f**, $T_2$ at the same fields. Filled symbols and solid lines include $T_1$ = 6 ms; open symbols omit it. The dotted line is the $^{13}$C limit and the dashed line $2T_1$. A polarized bath means $p$ = 1 for all orientations; "off-axis unpolarized" has $p$ = 1 on axis and 1/3 off axis. At 25 G the P1 echo is computed with the nitrogen nucleus explicit (Methods).

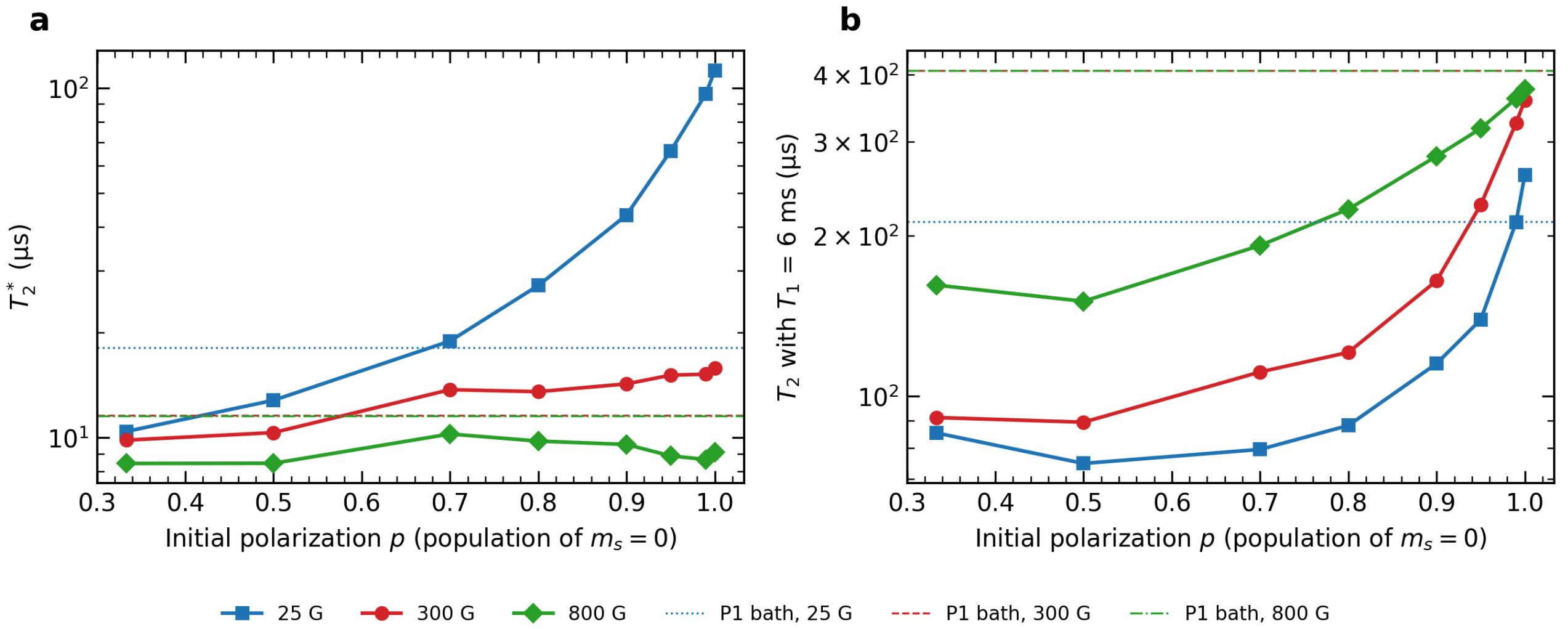

**Fig. 4 | Dependence on the initial bath polarization at $10^{17}$ cm$^{-3}$. a**, $T_2^*$ of a $^{15}NV^-$ probe against the population $p$ of $m_s = 0$ at 25 G (squares), 300 G (circles) and 800 G (diamonds). Horizontal lines are the P1 bath at each field (the 300 G and 800 G lines coincide). **b**, $T_2$ including $T_1 = 6$ ms. The polarization decays during the sequence, and $p$ is its initial value. $T_2$ rises steeply only above $p \approx 0.9$ and, at a given $p$, is longer at higher field, where the induced moments of the off-axis centres detune the minority flip-flops. The slightly higher value at $p = 1/3$ than at $p = 0.5$ ($\lesssim 13$ %) is within the spread of the configuration average and is not interpreted.

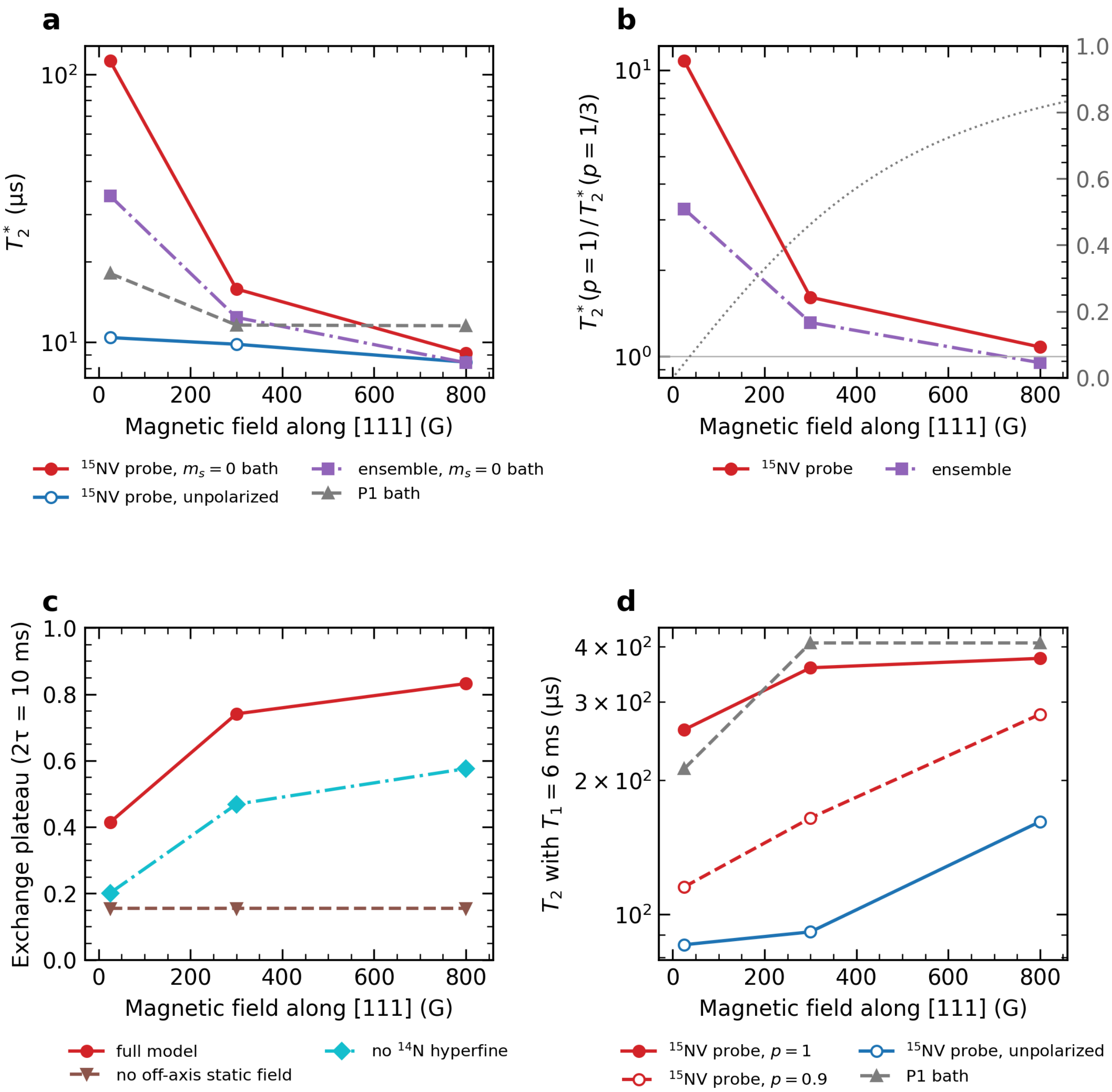

**Fig. 5 | Field dependence at $10^{17}$ cm$^{-3}$.** **a**, $T_2^*$ of a $^{15}$NV$^-$ probe in a fully polarized and an unpolarized NV$^-$ bath, of an ensemble measurement in a polarized bath, and of the P1 bath. **b**, Gain in $T_2^*$ from full polarization for the $^{15}$NV$^-$ probe and for an ensemble measurement, with the magnitude of the off-axis $m_s = 0$ moment $|\langle S_z\rangle|$ (dotted, right axis). **c**, Echo plateau of the exchange factor of a $^{14}$NV$^-$ probe in a fully polarized bath at $2\tau = 10$ ms, without $T_1$: full model, without the static field of the off-axis centres, and without the $^{14}$N hyperfine. **d**, $T_2$ including $T_1 = 6$ ms for a $^{15}$NV$^-$ probe in a bath with $p = 1$, $p = 0.9$ and $p = 1/3$, and for the P1 bath. The same off-axis moment that suppresses the $T_2^*$ gain localizes the exchange and slows the minority flip-flops.

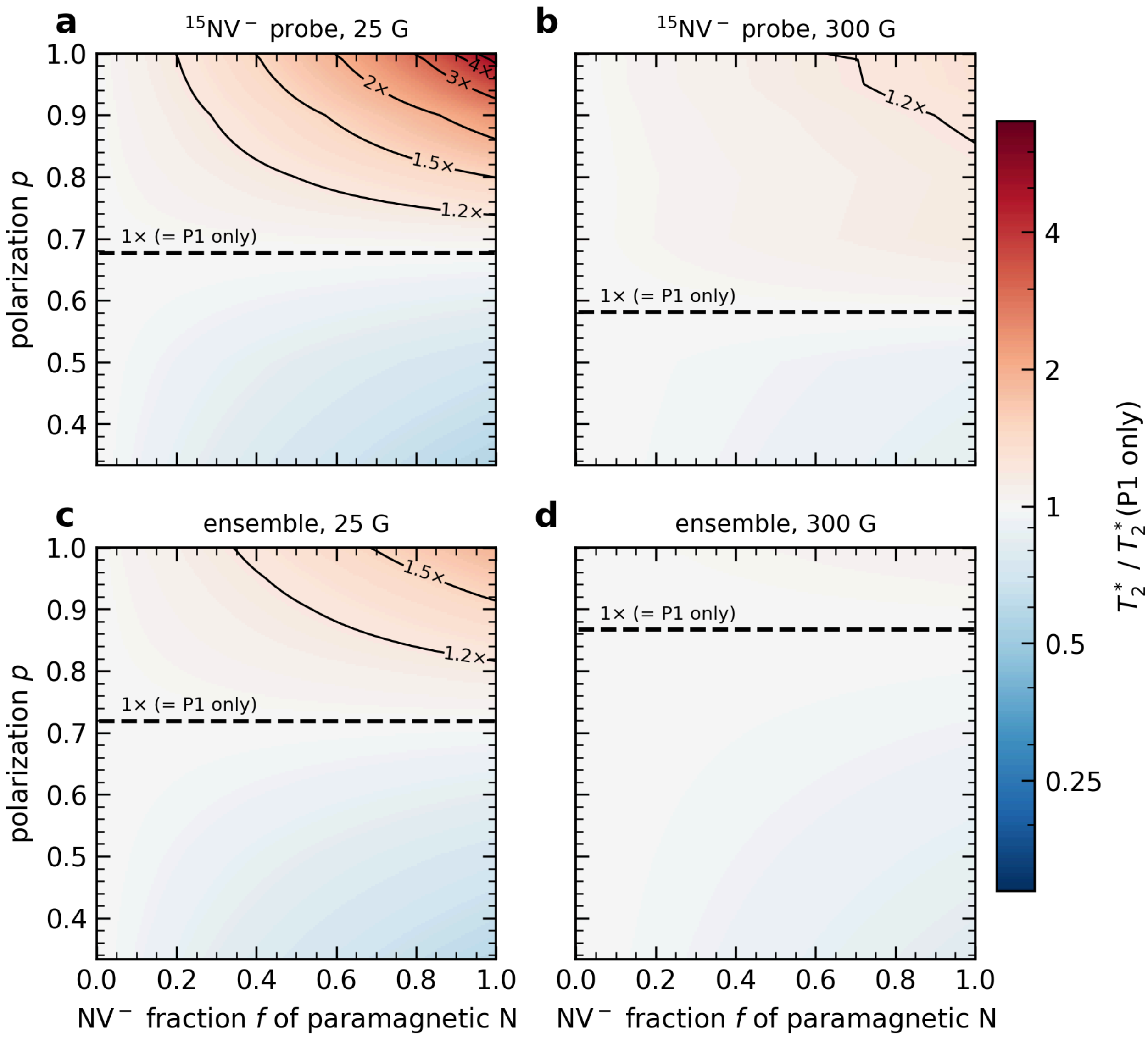

**Fig. 6 | Dephasing time against conversion and polarization at $10^{17}$ cm$^{-3}$.** Ratio of $T_2^*$ to that in a P1-only bath of the same total density, when a fraction $f$ of the paramagnetic nitrogen is NV$^-$ with polarization $p$ and the rest is P1. **a**,**b**, separately addressed $^{15}$NV$^-$ probe at 25 G (**a**) and 300 G (**b**). **c**,**d**, ensemble measurement driving all NV$^-$ of one orientation, at 25 G (**c**) and 300 G (**d**). Solid lines, contours of the ratio; dashed line, ratio 1, where the NV$^-$ bath alone dephases as fast as P1 ($p$ = 0.68, 0.58, 0.72 and 0.87 in **a**–**d**). Dephasing rates of the two baths are added (Methods).

## Tables

**Table 1 | Dephasing time $T_2^*$ (μs) at $10^{17}$ cm$^{-3}$ (0.57 ppm).** Single-probe values are from the exact solver (static fields and exchange); [a] static-field model. The $^{15}$NV$^-$ fraction is 1 % for the ensemble terms of the $^{15}$NV$^-$ probe.

| bath and probe | 25 G | 300 G | 800 G |
|---|---|---|---|
| P1 bath | 18.1 | 11.6 | 11.5 |
| $NV^-$ bath, unpolarized | 10.1 | 9.3 | 7.9 |
| $NV^-$ bath, $p = 1$, $^{15}NV^-$ probe | 112 | 15.8 | 9.1 |
| $NV^-$ bath, $p = 0.9$, $^{15}NV^-$ probe | 43.3 | 14.3 | 9.6 |
| $NV^-$ bath, on-axis $p = 1$, off-axis unpolarized[a] | 13.8 | 11.5 | 10.7 |
| $NV^-$ bath, $p = 1$, $^{14}NV^-$ probe (exchange) | 66 | 14.0 | 8.6 |
| ensemble, unpolarized bath | 10.7 | 9.4 | 8.8 |
| ensemble, polarized bath | 35.2 | 12.4 | 8.4 |
| $^{13}C$ (50 ppm) only | 405 | 428 | 431 |

**Table 2 | Hahn-echo time $T_2$ at $10^{17}$ cm$^{-3}$ (0.57 ppm).** Values include room-temperature $T_1$ = 6 ms; values without $T_1$ are in parentheses. [b] Echo plateau of the exchange factor at $2\tau = 10$ ms without $T_1$, where no 1/e time exists. "—": not applicable.

| bath and probe | 25 G | 300 G | 800 G |
|---|---|---|---|
| P1 bath | 212 μs (216 μs) | 408 μs (416 μs) | 408 μs (416 μs) |
| $NV^-$ bath, unpolarized, $^{15}NV^-$ probe | 85 μs (87 μs) | 91 μs (94 μs) | 162 μs (187 μs) |
| $NV^-$ bath, $p = 1$, $^{15}NV^-$ probe | 260 μs (9.9 ms) | 358 μs (9.9 ms) | 376 μs (9.9 ms) |
| $NV^-$ bath, $p = 0.9$, $^{15}NV^-$ probe | 115 μs (121 μs) | 165 μs (208 μs) | 282 μs (663 μs) |
| $NV^-$ bath, on-axis $p = 1$, off-axis unpolarized | — (76 μs) | — (76 μs) | — (120 μs) |
| $NV^-$ bath, $p = 1$, $^{14}NV^-$ probe (exchange) | 0.41[b] | 0.74[b] | 0.83[b] |
| ensemble, unpolarized bath | 35 μs (35 μs) | 33 μs (34 μs) | 40 μs (40 μs) |
| ensemble, polarized bath | 49 μs (50 μs) | 49 μs (50 μs) | 49 μs (50 μs) |
| $^{13}C$ (50 ppm) only | 197 ms | 200 ms | 195 ms |

# Supplementary Information

**Low-field dephasing time of an optically polarized $NV^-$ spin bath beyond the P1 limit in diamond: a cluster-correlation-expansion and exact spin-bath study**

Chikara Shinei

All calculations use a field along the NV [111] axis of 25, 300 or 800 G and 50 ppm $^{13}C$, the residual level of $^{12}C$-enriched crystals (main-text ref. 23). $NV^-$ bath centres are randomly oriented. Coherence times are the 1/e points of the simulated, configuration-averaged complex coherence; no lifetime is obtained by fitting. Reference numbers are those of the main text.

---

## Supplementary Note 1. Validation of the exact solver and the laboratory-frame pulse

The exact excitation-number-conserving solver (main text, Methods) was compared with full-cluster generalized CCE in PyCCE for a probe $NV^-$ and two on-axis bath $NV^-$ centres in $m_s = 0$. Two cases were tested: both bath centres resonant with the probe ($m_I = 0$), and one detuned by the $^{14}N$ hyperfine ($m_I = +1$). With the full cluster, generalized CCE is an exact calculation for this system. The two agree to within $3 \times 10^{-3}$ for the FID and the Hahn echo (Supplementary Fig. 1). The residual difference comes from non-secular dipolar terms, which the exact solver omits.

The agreement requires the time grid to be an integer multiple of the probe period $1/f_0$ ($f_0$ = 2029.05 MHz at 300 G). On an arbitrary grid the echo from generalized CCE differs from the exact result by up to 0.77, while the FID is unaffected. PyCCE applies an ideal pulse as an instantaneous rotation in the laboratory frame. When the probe exchanges resonantly with the bath, the echo has two pathways. In one the excitation stays on the probe; in the other it hops to a bath centre and is brought back by the second free evolution. The laboratory-frame pulse adds a phase $\exp(2i\omega_0\tau)$ between them. A real microwave pulse has a fixed phase in the frame rotating at the drive frequency and produces no such phase. On a commensurate grid the laboratory-frame pulse is identical to a rotating-frame pulse. This applies to any generalized-CCE calculation in which the qubit is resonant with part of its bath.

## Supplementary Note 2. Why generalized CCE is not the main method

Two failures of generalized CCE were found for these electron-spin baths. First, the second-order expansion diverges, for polarized and unpolarized $NV^-$ baths alike. The cluster contributions are ratios of subcluster coherences. When a resonant pair swaps its excitation, the subcluster coherence passes through zero.

Second, generalized CCE with sampled bath states needs the same large cluster cutoff as conventional CCE (Supplementary Note 7), which makes the second-order expansion with its divergences impractical for routine use.

The three calculations used instead are each exact for the effect they describe. Conventional CCE with sampled pure bath states and the mean field treats bath flip-flops at any polarization. The excitation-number solver treats probe–bath exchange to all orders. The Markov model treats $T_1$ jumps for independent centres.

## Supplementary Note 3. Localization of probe–bath exchange

Supplementary Fig. 2 shows the exchange factor of a $^{14}NV^-$ probe in a fully polarized bath at $10^{17}$ cm$^{-3}$, together with two controls. In the first control the static field of the off-axis centres is removed. In the second, the $^{14}N$ hyperfine is removed, so that all on-axis centres are resonant with the probe. The plateau of the full model at $2\tau$ = 10 ms rises with field, from 0.41 at 25 G to 0.74 at 300 G and 0.83 at 800 G. Removing the off-axis field lowers it to 0.16 at every field. Removing the hyperfine lowers it to 0.20, 0.47 and 0.58.

The plateau reflects localization of the single excitation on a dilute random lattice. The hopping between on-axis centres is $t = -\mathcal{J}_0(1 - 3\cos^2\theta)/(2r^3)$. At 300 G and 800 G the static disorder from the off-axis moments ($\langle S_z \rangle = -0.46$ and $-0.81$, three quarters of the bath) is several times $t$ at the typical spacing of the resonant sublattice (one centre in twelve). At 25 G it is an order of magnitude smaller. The plateau is nearly independent of concentration, because both scale as $1/r^3$. Supplementary Tables 1–3 list the results.

## Supplementary Note 4. $^{13}C$ ESEEM at 25 G

At 25 G the $^{13}C$ Larmor frequency is 26.8 kHz, and the $^{13}C$ factor of the Hahn echo shows electron-spin-echo envelope modulation (Supplementary Fig. 3a). At 50 ppm the modulation depth of the ensemble average is at most 8 %. Individual configurations with a $^{13}C$ close to the probe dip to about 0.3. The revivals return to about 1 at $2\tau = 2k/\nu_L$, and the long-time decay coincides with that at 300 G: $T_2(^{13}C)$ = 197 ms at 25 G against 200 ms at 300 G (Supplementary Fig. 3b). In the echo, the $^{13}C$ factor is replaced by its upper envelope through the revivals before it is multiplied with the electron-bath factors. The logarithmic time grid does not always sample a revival peak, but with an 8 % depth the effect on $T_2$ is negligible.

## Supplementary Note 5. The three effects of $T_1$

Supplementary Fig. 4 decomposes the echo of a $^{15}NV^-$ probe in a fully polarized bath at $10^{17}$ cm$^{-3}$ into its factors. The bath $T_1$ jumps alone decay with a 1/e time of 0.50, 0.47 and 0.42 ms at 25, 300 and 800 G. Their weak field dependence follows the static field distribution: the off-axis eigenstates carry larger moments at high field, so each jump moves a larger field. The flip-flop factor, evaluated at the polarization averaged over the sequence, decays with 0.30, 0.60 and 0.97 ms. Relaxation turns $m_s = 0$ centres into minority $m_s = \pm1$ centres that flip-flop, and at low field nothing detunes them. Together the two give 0.27, 0.37 and 0.39 ms; the probe $T_1$ factor and the instantaneous diffusion within the 1 % $^{15}NV^-$ group act near 10 ms. The exchange factor of the $^{15}NV^-$ probe stays above 0.95.

The jump-limited time scales as $n^{-1/2}$ (5.4, 1.5, 0.47 and 0.15 ms from $10^{15}$ to $10^{18}$ cm$^{-3}$ at 300 G for $p_0 = 1$). This scaling is that of sudden-jump spectral diffusion in a Lorentzian static field distribution (main-text ref. 18).

The jumps of a polarized bath are slightly faster than those of an unpolarized one (0.47 against 0.57 ms at $10^{17}$ cm$^{-3}$). Centres starting in $m_s = 0$ can jump to either $m_s = \pm1$, at a total rate 2Ω.

## Supplementary Note 6. The P1 dephasing slope

The static-field model gives $T_2^*\cdot$[P1] = 6.7–7.1 μs·ppm at 300 G from $10^{15}$ to $10^{18}$ cm$^{-3}$, and 6.5 μs·ppm at 800 G at $10^{17}$ cm$^{-3}$. This matches the analytic width of a dilute dipolar bath of $S = 1/2$ spins, 6.86 μs·ppm (main-text ref. 16). Second-order conventional CCE, which includes P1 flip-flops during the FID, changes the value by only 2 % (6.08 against 6.18 μs·ppm on a 192-configuration test set). Motional narrowing is therefore negligible.

The experimental slope is 9.6 ± 0.9 μs·ppm (main-text ref. 3). The 30 % difference most likely reflects how the P1 concentration is determined in experiment rather than missing physics. A nitrogen concentration measured by secondary-ion mass spectrometry includes non-paramagnetic nitrogen. The same static-field theory is used for every bath, so the ratios reported in the main text do not depend on this offset.

At 25 G the P1 slope is longer, 10.3 μs·ppm at $10^{17}$ cm$^{-3}$. Electron–$^{14}N$ mixing reduces the electron moment of four of the six P1 eigenstates ($|\langle \mathbf{S} \rangle| = 0.38$ and 0.04 for two pairs).

Averaging the magnitude $|L|$ over configurations before taking the 1/e point, instead of averaging the complex coherence, lengthens $T_2^*$ by about 1.5 times. Only the complex average corresponds to an ensemble signal.

## Supplementary Note 7. Convergence of the bath flip-flop calculation

The echo time set by bath flip-flops depends on two numerical choices: how the bath state enters, and the cutoff $r_{dip}$ below which two bath spins form a cluster. Supplementary Table 4 compares a fully mixed bath (density matrices, or no state for P1), as in the companion study,[5] with pure product states sampled from the eigenstates of each centre, for which PyCCE adds the static mean field of the spins outside each cluster. The test uses 300 G, $10^{17}$ cm$^{-3}$ (P1 also $10^{16}$ cm$^{-3}$), 6–8 configurations and, for sampled states, four samples per configuration.

**Supplementary Table 4 | $T_2 n$ (μs·ppm) of the bath flip-flop factor against the cluster cutoff $r_{dip}$ in units of the mean spacing $n^{-1/3}$.** “frozen”: the coherence stays above 1/e.

| bath | state | 1.2 | 2 | 3 | 4 | 5 |
|---|---|---|---|---|---|---|
| P1 | mixed | 85 | 69 | 65 | 63 | 66 |
| P1 | sampled + mean field | frozen | 295 | 214 | 252 | 282 |
| $NV^-$, $p = 1/3$ | mixed | 63 | 44 | 38 | 37 | — |
| $NV^-$, $p = 1/3$ | sampled + mean field | frozen | 83 | 62 | 57 | — |
| $NV^-$, $p = 0.9$ | mixed | plateau | 142 | 109 | 98 | — |
| $NV^-$, $p = 0.9$ | sampled + mean field | frozen | 204 | 132 | 137 | — |

Two errors appear. With a mixed bath, the mean field of the outer spins vanishes, close pairs flip-flop on resonance, and the echo time converges to a value 1.5–4 times too short. With sampled states, a cutoff near one spacing leaves the close pairs that are detuned by the mean field as the only clusters, and the bath appears frozen. With the mean field and a cutoff of 3–5 spacings, the P1 value agrees with the converged 236–286 μs·ppm of Park et al. (main-text ref. 24). The production runs use sampled states, a cutoff of 4 spacings and a bath radius of 5.5 spacings. $T_2 n$ agreed to 0.1 % between $10^{16}$ and $10^{17}$ $cm^{-3}$, so the flip-flop factor is computed at $10^{17}$ $cm^{-3}$ and rescaled in time to other densities.

An earlier version of this calculation used a mixed bath with a cutoff of 1.2 spacings. It gave the unpolarized $NV^-$ bath almost the same $T_2$ (by a cancellation of the two errors), but a P1 $T_2$ 3.2 times too short and a spurious plateau for partially polarized $NV^-$ baths.

---

## Supplementary Note 8. Field dependence of $T_2^*$ and the mean moment along the field

For a dilute bath of static dipoles the free-induction decay is Lorentzian with a width proportional to the density and to the mean moment of the bath centres along the field, $\langle|\langle S_z\rangle|\rangle$ (main-text ref. 16). Supplementary Table 5 lists this mean for each bath, computed from the exact eigenstates of each centre (P1: electron and $^{14}N$ with the full hyperfine and quadrupole tensors, four Jahn–Teller axes; $NV^-$: $D$ = 2.87 GHz, four orientations), and compares the field dependence it predicts with the calculated $T_2^*$ of Table 1.

**Supplementary Table 5 | Mean moment along the field and the field dependence of $T_2^*$ at $10^{17}$ $cm^{-3}$.** Ratios are normalized to 300 G. Polarized: $p$ = 1, $^{15}NV^-$ probe.

| bath | $\langle\|\langle S_z\rangle\|\rangle$ at 25 / 300 / 800 G | $T_2^*$ ratio, calculated | $1/\langle\|\langle S_z\rangle\|\rangle$ ratio |
|---|---|---|---|
| P1 | 0.30 / 0.50 / 0.50 | 1.56 / 1 / 0.99 | 1.65 / 1 / 0.99 |
| $NV^-$, unpolarized | 0.34 / 0.46 / 0.59 | 1.09 / 1 / 0.85 | 1.35 / 1 / 0.79 |
| $NV^-$, polarized | 0.033 / 0.35 / 0.61 | 7.1 / 1 / 0.58 | 10.7 / 1 / 0.57 |

The mean moment reproduces the field dependence of P1 within 6 % and the high-field side of both $NV^-$ baths. It overestimates the low-field gain of the $NV^-$ baths. The static model also includes the transverse components of $\langle \mathbf{S}\rangle$, which the projection ignores, and the polarized-bath values include the dispersive shifts of off-resonant exchange; both add dephasing that does not vanish with $\langle S_z\rangle$.

## Supplementary Figure legends

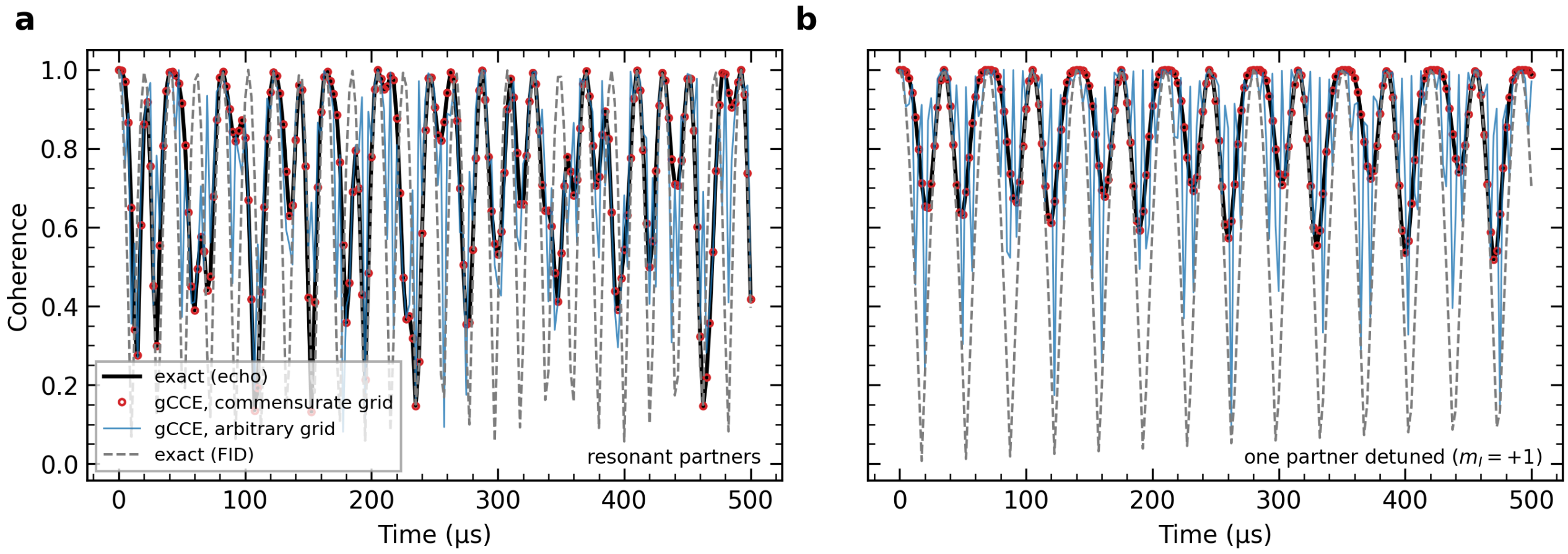

**Supplementary Fig. 1 | Validation of the exact solver.** Probe NV⁻ with two on-axis bath NV⁻ centres in $m_s = 0$ at 300 G. **a**, Both partners resonant ($m_I = 0$). **b**, One partner detuned by the $^{14}$N hyperfine ($m_I = +1$). Black: exact solver, Hahn echo. Red circles: generalized CCE on a time grid commensurate with the probe period. Blue: generalized CCE on an arbitrary grid, which carries the spurious laboratory-frame pulse phase. Grey dashed: exact FID.

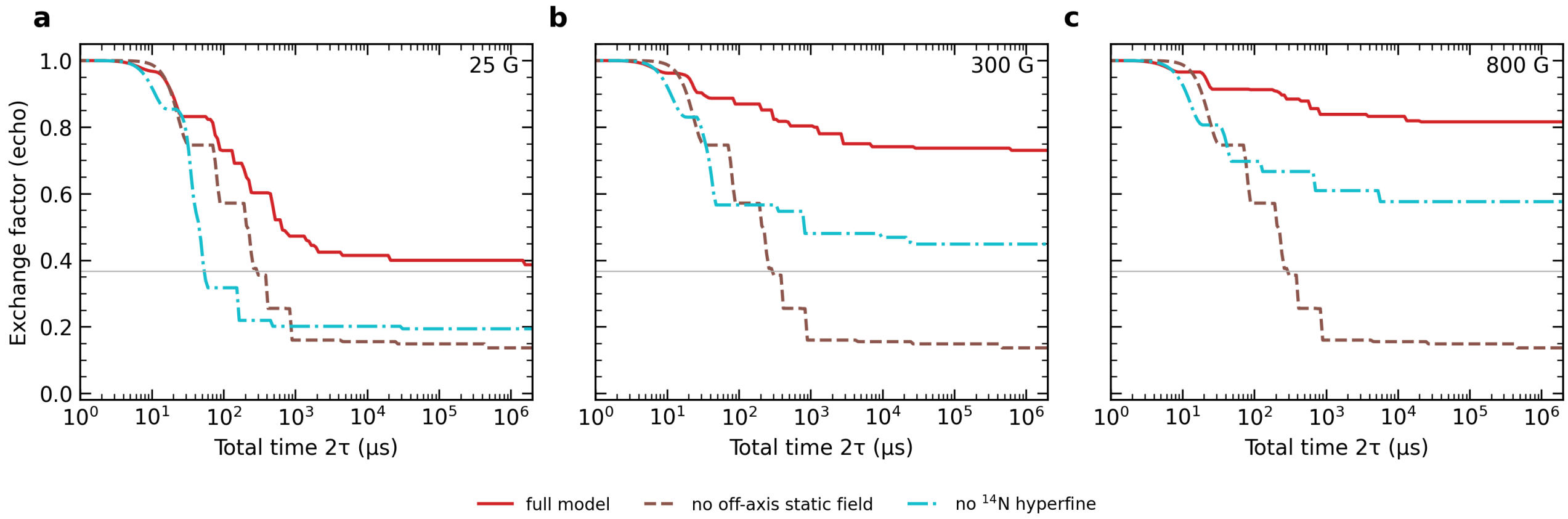

**Supplementary Fig. 2 | Exchange factor of a $^{14}NV^-$ probe in a fully polarized bath at $10^{17}$ cm$^{-3}$.** **a**, 25 G. **b**, 300 G. **c**, 800 G. Red: full model. Brown dashed: static field of the off-axis centres removed. Cyan dash-dotted: $^{14}N$ hyperfine removed. The full model averages 96 configurations, the controls 24.

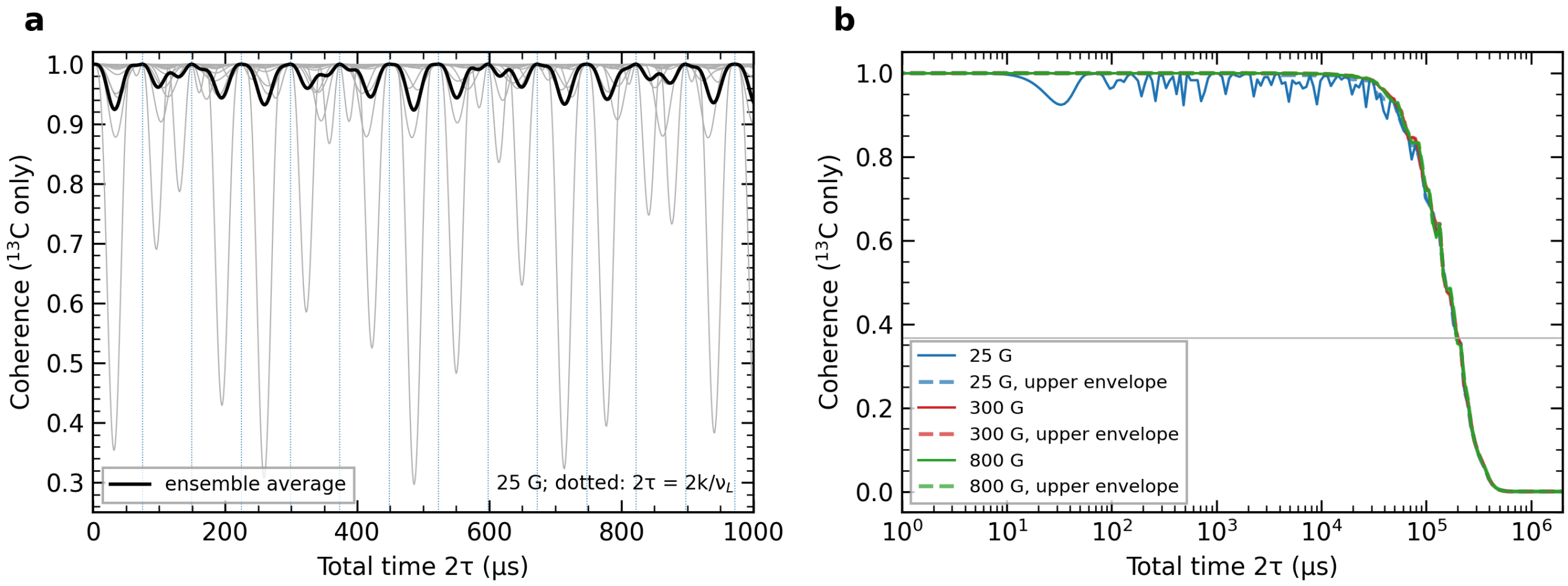

**Supplementary Fig. 3 | $^{13}C$ (50 ppm) contribution to the Hahn echo. a**, 25 G on a dense linear grid. Grey: individual configurations; black: ensemble average. Dotted vertical lines mark $2\tau = 2k/\nu_L$. **b**, The $^{13}C$ factor at 25, 300 and 800 G on the logarithmic grid of the main calculation, with its upper envelope (dashed).

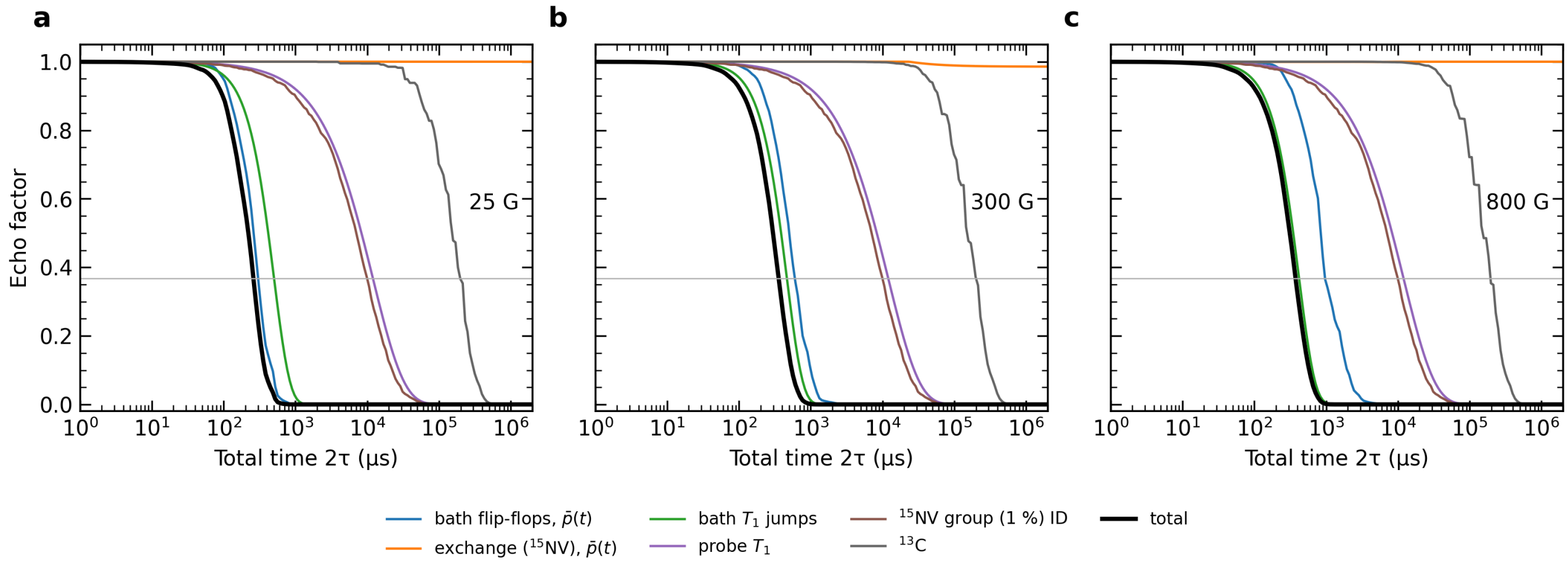

**Supplementary Fig. 4 | Factors of the Hahn echo of a $^{15}NV^-$ probe in a fully polarized bath at $10^{17}$ cm$^{-3}$ with $T_1$ = 6 ms.** **a**, 25 G. **b**, 300 G. **c**, 800 G. The bath $T_1$ jumps (green) and the flip-flop factor at the sequence-averaged polarization (blue) together set the total (black); the flip-flop factor dominates at 25 G.

---

## Supplementary Tables

**Supplementary Table 1 | $T_2^*$ (μs).** Single-probe values are from the exact solver with 480 configurations (static fields of frozen centres and probe–bath exchange); P1 and ensemble values from the static-field model with 1500 configurations; [a] static-field model. $^{13}$C (50 ppm) is included.

| bath and probe | field | $10^{15}$ cm$^{-3}$ | $10^{16}$ cm$^{-3}$ | $10^{17}$ cm$^{-3}$ | $10^{18}$ cm$^{-3}$ |
|---|---|---|---|---|---|
| P1 bath | 25 G | 338 | 137 | 18.1 | 1.91 |
| P1 bath | 300 G | 308 | 97.4 | 11.6 | 1.23 |
| P1 bath | 800 G | 307 | 96.6 | 11.5 | 1.22 |
| $NV^-$ bath, unpolarized ($^{14}NV^-$ probe) | 25 G | 281 | 87 | 10.1 | 1.05 |
| $NV^-$ bath, unpolarized ($^{14}NV^-$ probe) | 300 G | 287 | 73.6 | 9.33 | 0.861 |
| $NV^-$ bath, unpolarized ($^{14}NV^-$ probe) | 800 G | 283 | 70.5 | 7.91 | 0.776 |
| $NV^-$ bath, $p$ = 0.9 ($^{14}NV^-$ probe) | 25 G | 369 | 216 | 36.5 | 3.94 |
| $NV^-$ bath, $p$ = 0.9 ($^{14}NV^-$ probe) | 300 G | 308 | 102 | 13.1 | 1.18 |
| $NV^-$ bath, $p$ = 0.9 ($^{14}NV^-$ probe) | 800 G | 288 | 69.5 | 8.9 | 0.846 |
| $NV^-$ bath, $p$ = 1 ($^{14}NV^-$ probe) | 25 G | 393 | 277 | 66.3 | 7.6 |
| $NV^-$ bath, $p$ = 1 ($^{14}NV^-$ probe) | 300 G | 318 | 118 | 14 | 1.56 |
| $NV^-$ bath, $p$ = 1 ($^{14}NV^-$ probe) | 800 G | 288 | 74.8 | 8.59 | 0.937 |
| $NV^-$ bath, unpolarized ($^{15}NV^-$ probe) | 25 G | 284 | 87.9 | 10.4 | 1.06 |
| $NV^-$ bath, unpolarized ($^{15}NV^-$ probe) | 300 G | 289 | 74.5 | 9.84 | 0.865 |
| $NV^-$ bath, unpolarized ($^{15}NV^-$ probe) | 800 G | 284 | 71.1 | 8.44 | 0.786 |
| $NV^-$ bath, $p$ = 0.9 ($^{15}NV^-$ probe) | 25 G | 385 | 236 | 43.3 | 4.92 |
| $NV^-$ bath, $p$ = 0.9 ($^{15}NV^-$ probe) | 300 G | 315 | 107 | 14.3 | 1.26 |
| $NV^-$ bath, $p$ = 0.9 ($^{15}NV^-$ probe) | 800 G | 292 | 71.2 | 9.56 | 0.9 |
| $NV^-$ bath, $p$ = 1 ($^{15}NV^-$ probe) | 25 G | 400 | 330 | 112 | 13.6 |
| $NV^-$ bath, $p$ = 1 ($^{15}NV^-$ probe) | 300 G | 327 | 124 | 15.8 | 1.61 |
| $NV^-$ bath, $p$ = 1 ($^{15}NV^-$ probe) | 800 G | 294 | 77.7 | 9.1 | 0.987 |
| on-axis $p$ = 1, off-axis unpolarized[a] | 25 G | 319 | 104 | 13.8 | 1.35 |
| on-axis $p$ = 1, off-axis unpolarized[a] | 300 G | 306 | 91.8 | 11.5 | 1.13 |
| on-axis $p$ = 1, off-axis unpolarized[a] | 800 G | 303 | 85.7 | 10.7 | 1.04 |
| ensemble, unpolarized bath | 25 G | 297 | 84.9 | 10.7 | 1.07 |
| ensemble, unpolarized bath | 300 G | 294 | 75.3 | 9.44 | 0.914 |
| [illegible] | [illegible] | [illegible] | [illegible] | [illegible] | [illegible] |

**Supplementary Table 2 | $T_2$ (μs) with room-temperature $T_1$ = 6 ms.** Bath flip-flops, probe–bath exchange, bath $T_1$ jumps, probe $T_1$, the $^{15}NV^-$ group (1 %) and $^{13}C$ (50 ppm) are included (Methods). P1 includes only the probe $T_1$; the P1 echo is not computed at 25 G.

| bath and probe | field | $10^{15}$ cm$^{-3}$ | $10^{16}$ cm$^{-3}$ | $10^{17}$ cm$^{-3}$ | $10^{18}$ cm$^{-3}$ |
|---|---|---|---|---|---|
| P1 bath | 25 G | 8770 | 1920 | 212 | 21.6 |
| P1 bath | 300 G | 8570 | 3390 | 407 | 41.6 |
| P1 bath | 800 G | 8570 | 3390 | 407 | 41.6 |
| $NV^-$ bath, unpolarized ($^{14}NV^-$ probe) | 25 G | 3890 | 737 | 83.8 | 8.52 |
| $NV^-$ bath, unpolarized ($^{14}NV^-$ probe) | 300 G | 3820 | 750 | 89 | 9.01 |
| $NV^-$ bath, unpolarized ($^{14}NV^-$ probe) | 800 G | 3890 | 985 | 158 | 17.2 |
| $NV^-$ bath, $p$ = 0.9 ($^{15}NV^-$ probe) | 25 G | 3870 | 855 | 115 | 11.9 |
| $NV^-$ bath, $p$ = 0.9 ($^{15}NV^-$ probe) | 300 G | 3830 | 994 | 165 | 19.8 |
| $NV^-$ bath, $p$ = 0.9 ($^{15}NV^-$ probe) | 800 G | 3760 | 1150 | 282 | 49.5 |
| $NV^-$ bath, $p$ = 1 ($^{15}NV^-$ probe) | 25 G | 3940 | 998 | 260 | 74 |
| $NV^-$ bath, $p$ = 1 ($^{15}NV^-$ probe) | 300 G | 3810 | 1160 | 358 | 110 |
| $NV^-$ bath, $p$ = 1 ($^{15}NV^-$ probe) | 800 G | 3730 | 1160 | 376 | 115 |
| ensemble, unpolarized bath | 25 G | 2410 | 320 | 34.9 | 3.43 |
| ensemble, unpolarized bath | 300 G | 2340 | 308 | 33.3 | 3.27 |
| ensemble, unpolarized bath | 800 G | 2430 | 361 | 39.5 | 3.83 |
| ensemble, polarized bath | 25 G | 2570 | 426 | 49.1 | 4.81 |
| ensemble, polarized bath | 300 G | 2490 | 421 | 49 | 4.81 |
| ensemble, polarized bath | 800 G | 2430 | 417 | 48.9 | 4.81 |

**Supplementary Table 3 | $T_2$ (μs) without $T_1$.** As Supplementary Table 2 but without any longitudinal relaxation.

| bath and probe | field | $10^{15}$ cm$^{-3}$ | $10^{16}$ cm$^{-3}$ | $10^{17}$ cm$^{-3}$ | $10^{18}$ cm$^{-3}$ |
|---|---|---|---|---|---|
| P1 bath | 25 G | $2.1 \times 10^4$ | 2150 | 216 | 21.6 |
| P1 bath | 300 G | $4 \times 10^4$ | 4160 | 415 | 41.7 |
| P1 bath | 800 G | $4 \times 10^4$ | 4160 | 415 | 41.7 |
| $NV^-$ bath, unpolarized ($^{14}NV^-$ probe) | 25 G | 8320 | 839 | 85.3 | 8.54 |
| $NV^-$ bath, unpolarized ($^{14}NV^-$ probe) | 300 G | 9030 | 908 | 91.3 | 9.04 |
| $NV^-$ bath, unpolarized ($^{14}NV^-$ probe) | 800 G | $1.8 \times 10^4$ | 1780 | 180 | 17.5 |
| $NV^-$ bath, $p = 0.9$ ($^{15}NV^-$ probe) | 25 G | $1.2 \times 10^4$ | 1210 | 121 | 12 |
| $NV^-$ bath, $p = 0.9$ ($^{15}NV^-$ probe) | 300 G | $2 \times 10^4$ | 2090 | 208 | 20.5 |
| $NV^-$ bath, $p = 0.9$ ($^{15}NV^-$ probe) | 800 G | $5.6 \times 10^4$ | 6630 | 663 | 63.7 |
| $NV^-$ bath, $p = 1$ ($^{15}NV^-$ probe) | 25 G | $1.8 \times 10^5$ | $7.9 \times 10^4$ | 9910 | 934 |
| $NV^-$ bath, $p = 1$ ($^{15}NV^-$ probe) | 300 G | $1.8 \times 10^5$ | $8 \times 10^4$ | 9950 | 904 |
| $NV^-$ bath, $p = 1$ ($^{15}NV^-$ probe) | 800 G | $1.8 \times 10^5$ | $8 \times 10^4$ | 9940 | 845 |
| ensemble, unpolarized bath | 25 G | 3380 | 340 | 35.1 | 3.44 |
| ensemble, unpolarized bath | 300 G | 3230 | 323 | 33.5 | 3.27 |
| ensemble, unpolarized bath | 800 G | 3860 | 390 | 39.8 | 3.84 |
| ensemble, polarized bath | 25 G | 4770 | 479 | 49.8 | 4.82 |
| ensemble, polarized bath | 300 G | 4790 | 479 | 49.8 | 4.82 |
| ensemble, polarized bath | 800 G | 4790 | 479 | 49.8 | 4.82 |